\documentstyle[12pt]{article}
\def\hybrid{\topmargin -20pt	\oddsidemargin 0pt
	\headheight 0pt	\headsep 0pt
	\textwidth 6.25in	
	\textheight 9.5in	
	\marginparwidth .875in
	\parskip 5pt plus 1pt	\jot = 1.5ex}

\hybrid
\newskip\humongous \humongous=0pt plus 1000pt minus 1000pt
\def\caja{\mathsurround=0pt}
\def\eqalign#1{\,\vcenter{\openup1\jot \caja
	\ialign{\strut \hfil$\displaystyle{##}$&$
	\displaystyle{{}##}$\hfil\crcr#1\crcr}}\,}
\newif\ifdtup

\relax

\def\e{\epsilon}
\def\s{\sigma}

\def\be{\begin{equation}}
\def\ee{\end{equation}}
\def\ba{\begin{eqnarray}}
\def\ea{\end{eqnarray}}
\def\d{\partial}
\def\xb{{\bar x}}
\def\db{\bar{\partial}}

\def\s{\sigma}

\def\p{\psi}
\def\px{{\vec p}}
\def\e{\epsilon}

\def\Kuu{K_{u\bar u}}
\def\Kvv{K_{v\bar v}}
\def\U{\log\Kuu}

\def\rr{{\rm regular}}

\def\ec{\hat E^{c}_{2}}
\def\pp{\bar\p}
\def\j{\tilde J}

\begin{document}
\renewcommand{\theequation}{\thesection.\arabic{equation}}
\newcommand{\beq}{\begin{equation}}
\newcommand{\eeq}[1]{\label{#1}\end{equation}}
\newcommand{\ber}{\begin{eqnarray}}
\newcommand{\eer}[1]{\label{#1}\end{eqnarray}}
\begin{titlepage}
\begin{center}

\hfill CERN-TH.7218/94\\
\hfill HUB-IEP-94/5\\
\hfill LPTENS-94/10\\
\hfill hep-th/9404114\\

\vskip .7in

{\large \bf Superstring Gravitational Wave Backgrounds
with Spacetime Supersymmetry}
\vskip .6in

{\bf Elias Kiritsis and Costas Kounnas\footnote{On leave from Ecole
Normale Sup\'erieure, 24 rue Lhomond, F-75231, Paris, Cedex 05,
FRANCE.}}\\
\vskip
 .1in

{\em Theory Division, CERN, CH-1211\\
Geneva 23, SWITZERLAND} \footnote{e-mail addresses:
KIRITSIS@NXTH04.CERN.CH, KOUNNAS@NXTH04.CERN.CH}\\

\vskip .2in
and
\vskip .2in

{\bf Dieter L\"ust}
\vskip .1in

{\em  Humboldt Universit\"at zu Berlin\\
Institut f\"ur Physik\\
D-10099 Berlin, GERMANY}\footnote{e-mail address:
LUEST@QFT1.PHYSIK.HU-BERLIN.DE

}

\end{center}

\vskip .2in

\begin{center} {\bf ABSTRACT } \end{center}
\begin{quotation}\noindent
We analyse the stringy gravitational wave background based on the
current algebra $E^{c}_{2}$. We determine its exact spectrum and
construct the modular invariant vacuum energy. The corresponding N=1
extension is also constructed.
The algebra is again mapped to free bosons and fermions and we show
that this background has N=4 (N=2) unbroken spacetime supersymmetry
in
the type II (heterotic case).

\end{quotation}
\vskip 2.0cm
CERN-TH.7218/94 \\
April 1994\\
\end{titlepage}
\vfill
\eject
\def\baselinestretch{1.2}
\baselineskip 16 pt
\setcounter{equation}{0}

Gravitational wave-like backgrounds \cite{gw} have special features
which render them interesting testing grounds of our understanding of
stringy gravity. This  subject has been revived recently,
\cite{w}-\cite{r1,rest} with focus on current algebra generated
gravitational wave backgrounds.

In this paper, we will discuss an exact  classical solution to string
theory
found by Nappi and Witten \cite{w}, which can be interpreted as a
plane
gravitational (as well as $B_{\mu\nu}$) wave. It has large symmetry
(seven Killing symmetries) and the corresponding 2-d $\sigma$-model
is
the WZW model for a non semi-simple group, $E^{c}_{2}$, the central
extension of the 2-d Euclidean group.
This model is interesting because it provides us with an exact
classical solution to string theory which has at the same time:

$\bullet$ a simple and clear physical interpretation,

$\bullet$ a non-trivial spacetime,

$\bullet$ it can be solved,

$\bullet$ it has unusual features, in particular the spacetime is
nowhere
asymptotically flat. Thus, a priori, it is not obvious that a
sensible
scattering matrix exists in such a background.

In \cite{kk} we began solving the model. In particular, among other
things, we mapped the current algebra and the representation theory
to
(almost) free fields.
In this note we present a continuation of our effort to understand
the
physics
of the model.

We are interested in considering this background as an
exact classical solution to superstring theory.
Thus we will supersymmetrize the current algebra and we will again
map
the model to free bosons and fermions.
By studying the $\s$-model we will show
that the model has a $N=4$
worldsheet supersymmetry. The associated string
theory has a large unbroken spacetime supersymmetry.
For the type II string this is  N=4 spacetime supersymmetry.
In fact, this type of 4-d background turns out to
be a special case of the class of solutions found recently in
\cite{kkl}.
Moreover, we will explicitly construct the N=4 superconformal algebra
out of the (super) current algebra.

Some of the features though of the model like the potential spectrum
and and tree scattering of bosonic states have little difference
between the bosonic and the supersymmetric model. Thus we will find
the exact spectrum of the associated bosonic string theory
and we will compute the (modular invariant) vaccum energy.
There are two types of states in the theory, corresponding to
different kinds
of representations of the current algebra. We will qualitatively
describe their scattering.

We will start our discussion by presenting the current algebra
symmetry
of the background \cite{w}.
This is specified by the OPEs
$$J_{a}(z)J_{b}(w)={G_{ab}\over (z-w)^2}+{f_{ab}}^{c}{J_{c}(w)\over
(z-w)}+
\rr,\eqno(1)$$
where $(J_{1},J_{2},J_{3},J_{4})\sim(P_{1},P_{2},J,T)$, the $P_{i}$
generating the translations and $J$ being the generator of rotations
and $T$ being the central element.
The only non-zero structure constants are ${f_{31}}^{2}=1$
${f_{32}}^{1}=-1$ and ${f_{12}}^{4}=1$;
$G_{ab}$ is an invariant bilinear form (metric) of the algebra. $G$
is a symmetric matrix
and the Jacobi identities
constrain it so that $f_{abc}\equiv {f_{ab}}^{d}G_{dc}$ is completely
antisymmetric.
The most general invariant bilinear form for $E^{c}_{2}$ is
$$G=k\left(\matrix{1&0&0&0\cr 0&1&0&0\cr 0&0&b&1\cr
0&0&1&0\cr}\right).\eqno(2)$$
where we can assume without loss of generality that $k>0,b>0$
\cite{kk}.
For $\hat E^{c}_{2}$ there is a unique solution to the Master
Equation \cite{HK}, which has
all the
properties of the Affine Sugawara construction. It is given by:
$$L_{AS}={1\over 2k}\left(\matrix{1&0&0&0\cr 0&1&0&0\cr 0&0&0&1\cr
0&0&1&-b+{1\over k}\cr}\right)={1\over 2}G^{-1}+{1\over 2k^2}
\left(\matrix{0&0&0&0\cr 0&0&0&0\cr 0&0&0&0\cr
0&0&0&1\cr}\right),\eqno(3)$$

The $\s$-model realizing the current algebra above can be constructed
in a straightforward fashion, and provides the tool to study string
propagation.
Its action is \cite{w}
$$S={k\over 2\pi}\int d^{2}z\left[\d a_{i}\db a_{i}+\d u\db v+\db u\d
v
+b\d u\db u-\e^{ij}a_{i}\db a_{j}\d u\right]\eqno(4)$$

{}From now on we will set $b=0$ by an appropriate shift in $v$. As
was
shown in \cite{kk}, from the exact current algebra point of view,
once $b$ is finite
it can always rescaled away (assuming that the light-cone coordinates
are
non-compact).
We can also scale $k\to 1$ by appropriate scaling of
$a_{i},v$\footnote{
Effectively we are setting the parameter $Q$ of \cite{kk} to one.}.
In this coordinate system, the nature of the background is not
obvious, however
as we will show below, it is in this background that the free field
representation of the algebra described in \cite{kk} is manifest.
It should be also noted that in (4) the last term describes the
departure of the background from flat Minkowski space, and its
coupling can be made arbitrarily
small by a boost of the $u,v$ coordinates.
However, this does not imply that the perturbation is insignificant
since
the perturbing operator has bad IR behaviour.
It should also be mentioned here that the action (4) can be obtained
by an
$O(3,3,R)$ transformation of flat space, \cite{kk,kt}.

The coordinate system where the nature of the background is more
transparent is given by \cite{w}
$$a_{1}=x_{1}+\cos u x_{2}\;\;,\;\; a_{2}=\sin u x_{2}\;\;,\;\;
v\to v+{1\over 2}x_{1}x_{2}\sin u\eqno(5)$$
where the action (4) becomes
$$S={1\over 2\pi}\int d^2 z\left[\d x_{i}\db x_{i}+2\cos u\d x_{2}\db
x_{1}+\d u\db v+\db u\d v\right].\eqno(6)$$
In this form we can immediately identify the perturbation from flat
space
with a graviton+antisymmetric tensor mode given by $\cos u\d x_{2}\db
x_{1}$.
This is an operator with no IR problems, however its coupling is of
order one and cannot be rescaled at will.
One can however look at the structure of perturbation theory around
the flat background, and verify that indeed the current algebra
structure decribed in \cite{kk} indeed emerges.

In \cite{kk} we described a resolution of the $\ec$ current algebra
in terms
of free fields and studied its irreducible representations with
unitary base.
We will give here a brief summary of the results that we will need
and refer the reader to \cite{kk} for more details.
We will focus for the moment on one copy of the current algebra, and
return to
the full theory in due time.
We introduce four free fields, $x^{\mu}$ with
$$\langle x^{\mu}(z)x^{\nu}(w)\rangle=\eta^{\mu\nu}\log(z-w)\;\;,\;\;
\eta^{\mu\nu}\sim\left(\matrix{1&0&0&0\cr
0&-1&0&0\cr 0&0&-1&0\cr 0&0&0&-1\cr}\right)\eqno(7)$$
and define the light-cone $x^{\pm}=x^{0}\pm x^{3}$ and transverse
space
$x=x^{1}+ix^{2},\bar x=x^{1}-ix^{2}$ coordinates.
Then, the current algebra generators can be uniquely (up to Lorentz
transformations) written as
$$J={1\over 2}\d x^{+}\;\;\;,\;\;\;T=\d x^{-}\eqno(8)$$
$$P^{+}=P^{1}+iP^{2}=ie^{-ix^{-}}\d
x\;\;\;,\;\;\;P^{-}=P^{1}-iP^{2}=ie^{ix^{-}}\d \xb\eqno(9)$$
while the affine-Sugawara stress tensor (3) becomes the free field
stress
tensor
$$T_{AS}={1\over 2}\eta_{\mu\nu}\d x^{\mu}\d x^{\nu}.\eqno(10)$$

The current algebra representations can be built by starting with
unitary
representations of the $\ec$ algebra \cite{str} as a base and then
acting on them by the negative modes of the currents.
They fall into two classes, \cite{kk}:

(I) Representations with neither highest nor lowest weight
state.\footnote{
Except when $\vec p=0$ in which case the representations are one
dimensional.
}
The base is generated by vertex operators with $p_{-}=0$.
The representation is characterized by $p_{+}\in [0,1)$ and
transverse momentum $\px_{T}$.
The base operators are
$$V^{I}_{p_{+},n}=e^{i(p_{+}+n)x^{-}+i\px_{T} \cdot \vec
x}.\eqno(11)$$

(II) Representations with a highest weight state in the base.
Their conjugates have a lowest weight.
They have $p_{-}>0$ (negative for the conjugates).
The highest weight operator can be written as
$$V^{II}\sim e^{ip_{+}x^{-}+ip_{-}x^{+}}H_{p_{-}}\eqno(12)$$
where $H_{p_{-}}$ is the generating twist field in the transverse
space
corresponding to the transformation
$$x(e^{2\pi i}z)=e^{-4\pi ip_{-}}x(z)\;\;,\;\;\bar
x(e^{2\pi i}z)=e^{4\pi ip_{-}}\xb(z).\eqno(13)$$
The conformal weight of the operators at the base is
$\Delta=-2p_{+}p_{-}+p_{-}(1-2p_{-})$.

As we will see later on all the representations participate in the
modular invariant vacuum energy.
Thus we can easily describe the spectrum of physical states.
We will assume that the extra 22 dimensions are non-compact although
the compact case is as easy to handle.

$\bullet$ For type I states ($p_{-}=0$) we have the physical state
conditions
$L_{0}=\bar L_{0}=1$ which translate  to
$$N=\bar N\;\;\;{\rm and}\;\;\;\vec
p_{T}^2=2-2N\;\;\;N=0,1,2,\cdots\eqno(14)$$
where, $N$,$\bar N$ are the contribution of the transverse currents
or oscillators (from the left or right) and $\vec p_{T}$ is the 24-d
transverse momentum.
It is obvious that for $N=0$ we have $\vec p_{T}^2=2$ which is a
component
of the tachyon, while for $N=1$ we have $\vec p=0$ and we obtain some
modes
of the massless fields (graviton, antisymmetric tensor and dilaton).
which propagate along the wave.
For $N>1$ there are no physical states.

$\bullet$. For type II states the situation is like in usual string
theory,
it is just their dispersion relation that changes:
$$-4p_{+}p_{-}+2p_{-}(1-2p_{-})+\vec p_{T}^2=2-2N\eqno(15)$$
and $\vec p_{T}$ now stands for the 22 transverse dimensions.
Eq. (15) can also be written after $p_{-}\to p_{-}/4$, $p_{+}\to
p_{+}-p_{-}/4+1/2$ as
$$p_{+}p_{-}-\vec p_{T}^{2}=2(N-1)\eqno(16)$$
which is the flat space spectrum but in 24-d Minkowski space.
Thus for almost all states the wave reduces the effective
dimensionality by two. As we will see below, this happens because the
states are localized in the extra two dimensions.

Eventually we would lie to compute the partition function (vaccuum
energy).
In order to do this we can guess the modular invariant sowing of
representations by looking at the wavefunctions on the group.
Using the results of \cite{sf} these wavefunctions can be computed
and
and we easily deduce that we have the standard diagonal modular
invariant.

{}From these wavefunctions we can see the qualitative behaviour of
the fluctuations in this background.
Type I states are just plane waves. As for type II states,
if we go to the $x_{i}$ coordinates (5) where the wave nature is
manifest,
then we see that the states are localized where $x^{2}_{1}+x^{2}_{2}+
2x_{1}x_{2}\cos(u)=0$. So for fixed time the state is localized  on
the two
lines $x_1=\pm x_{2}$, $x_3=$constant where $u=t-x_{3}$.
This disturbance travels to the left in the $x_3$ direction with the
speed of light (in flat space).

We can now try to construct the vaccuum energy by putting together
the representations.
What can happen at most is the truncation phenomenon as in the
compact case.
In trying to built the modular invariant partition function we have
to overcome
the problem that the representations are infinite dimensional and
since all the states in the base have the same energy the partition
function diverges.
We will have to regulate this divergence and ensure that upon the
removal
of the regulator we will obtain a modular invariant answer.
We will start by computing the character formulae for the
representations above.
In fact what we are interested in is the so called signature
character which keeps track of the positive and negative norms.
We will not worry much about the type I representations (although
they are the easiest to deal with) since their contribution is of
measure zero.
For the type II representations we can easily compute the signature
character
$$\chi^{II}_{p_{+},p_{-}}(q,w)\equiv
Tr[(-1)^{\epsilon}q^{L_{0}-{c\over 24}}w^{J_{0}}]\eqno(17)$$
where $\epsilon =0,1$ for positive (respectively negative) norm
states.
If $2p_{-}\not=$integer\footnote{We can also compute the character
when $2p_{-}\in Z$ but this is not needed again for the partition
functions since it is of measure zero.} then the only non-trivial
null vector is the one at the
base responsible for the fact that we have a highest (or lowest)
weight representation. Thus the character is \cite{sf2,BK}
$$\chi^{II}_{p_{+},p_{-}}(q,w)={q^{-2p_{+}p_{-}+p_{-}(1-2p_{-})
-{1\over 6}}
w^{p_{+}}\over (1-w)\prod_{n=1}^{\infty}(1-q^{n}w)(1-q^{n}w^{-1})}=$$
$$=iq^{-2p_{+}p_{-}+p_{-}(1-2p_{-})-{1\over 12}}w^{p_{+}-{1\over 2}}
{\eta(q)\over \vartheta_{1}(v,q)}
\eqno(18)$$
where $\vartheta_{1}$ is the standard Jacobi $\vartheta$-function and
$w=e^{2\pi i v}$.
The infinity we mentioned before can be seen as $w\to 1$ as the pole
in the $\vartheta$-function.
In order to see how we must treat this infinity we will calculate
this partition function in the quantum mechanical case (that is
keeping
track only of the zero modes).
We will introduce a twisted version of the (Minkowski) amplitute for
the particle to go from $x$ to $x'$
in time $\tau$:
$$<x|x';\tau,v,\bar v>=<x|e^{i\tau H}e^{\zeta
J_{0}+\bar \zeta\bar
J_{0}}|x'>\eqno(19)$$
The quantum mechanical partition function can be obtained by setting
$x=x'$, $\zeta,\bar \zeta\to 0$ and integrating over $x$.
This last integral gives the overall volume of space that we will
drop
since we are interested in the vaccuum energy per unit volume.
The amplitude (19) in the coordinates $u,v$ and
$a_{1}=r\cos\theta,a_{2}=r\sin\theta$, can be
calculated with the result
$$<x|x';\tau,\zeta,\bar \zeta>\sim {1\over \tau^2}{(u-u'+\bar
\zeta-\zeta)\over \sin\left(
{u-u'+\bar \zeta-\zeta\over 2}\right)}\exp\left[i{(u-u'+\bar
\zeta-\zeta)^2\over 4\tau}-i
{(u-u'+\bar \zeta-\zeta)(v-v')\over 2\tau}-\right.$$
$$-\left.{i\over 8\tau}(u-u'+\bar \zeta-\zeta)
\cot\left({u-u'+\bar \zeta-\zeta\over
2}\right)\left(r^2+r'^2-2rr'{\cos\left(
\theta-\theta'-{u-u'+\zeta+\bar \zeta\over 2}\right)\over
\cos\left({u-u'+\bar \zeta-\zeta\over 2}\right)}\right)\right]$$
We can then perform the limits to obtain the $finite$ result
$$Z(\tau)=<x|x;\tau,\zeta=0,\bar \zeta=0>\sim {1\over
\tau^{2}}\eqno(20)$$
This result may seem surprising since we have only two continuous
components of the momentum, namely $p_{+}$ and $p_{-}$ but no
transverse
momentum.
However the result is not as surprising as it looks at first, and to
persuade the reader  to that we will provide with an analogous
situation where the answer is obvious.
Consider the case of a flat 2-d plane. The zero mode spectrum are the
usual plane waves, $e^{i\vec p\cdot \vec x}$, but we will work in the
rotational
basis, where the appropriate eigenfunctions corresponding to energy
$p^2$
are $e^{im\theta}J_{|m|}(pr)$.
In such a basis, we will have precisely the same problem in
calculating
the partition function as we had above. Namely, for a given $p$ there
are an infinite number of states (numbered by $m\in Z$) with the same
energy.
Thus the partition function seems to be infinite.
However here we know what to do: go to the (good) plane wave basis,
or
calculate the propagator and then take the points to coincide in
order
to get the partition function. This will also give the right result.
This is precisely the prescription we have applied above and we found
that
the quantum mechanical partition function of the gravitational wave
zero modes is precisely that of flat Minkowski space.
Having found the way to sum the zero mode spectrum, it is not
difficult to show
using (20) that the vaccum energy of the associated bosonic string
theory
is equal to the flat case, namely
$$F=\int{d\tau d\bar \tau\over Im\tau^2}(\sqrt{Im\tau}\eta\bar
\eta)^{-24}\eqno(21)$$
where we have added another 22 flat non-compact dimensions\footnote{
The extra 22 dimensions could be compactified with no additional
effort}.

We would add here a few comments about scattering. We hope to present
the full
picture in the future.
States corresponding to type I representations scatter among
themselves, this is already obvious from their vertex operator
expressions \cite{kk}.
This is dangerous however, since we might have trouble with
unitarity.
We will check here that in a 4-point amplitude of type I tachyons
only
physical states appear as intermediate states.
Remember that type I tachyons have $p_{-}=0$. Thus the four type I
tachyons are characterized  by their $p_{+}^{i}$ and $\vec
p_{T}^{i}$.
The 4-point amplitude is then given by the standard
$\delta$-functions
of $p_{+}$ and $\vec p_{T}$ multiplied by the Shapiro-Virasoro
amplitude
with one difference: instead of the invariants $p_{i}\cdot p_{j}$ we
now have them restricted to transverse space, $\vec p^{i}_{T}\cdot
\vec
p^{j}_{T}$.
This is similar to the Shapiro-Virasoro amplitude in 24 Euclidean
dimensions
(modulo the delta functions)
and its analytic structure is different.
It can be easily checked that instead of the infinite sequence of
poles
of the usual amplitude here we have just two poles, one corresponding
to intermediate on-shell tachyons and the other to intermediate
on-shell massless type I particles. This is in agreement with the
fusion rules of the current algebra.\footnote{Scattering for type I
states
for arbitrary plane waves has been considered in \cite{r1} with similar
results.}

The scattering of the type II states is more complicated and will be
dealt with elsewhere.

The bosonic string vacuum  as it stands is ill defined in the quantum
theory due to the presence of the tachyon in its spectum.
We will construct now the N=1 extension of this background, in order
to
make a superstring out of it.

In order to do this we  will add four free fermions (with Minkowski
signature),
$\psi_{1,2}$, $\psi_{J}$, $\psi_{T}$, normalized as
$$\psi_{a}(z)\psi_{b}(w)={G_{ab}\over (z-w)}+\rr\eqno(22)$$
The N=1 supercurrent is given by
$$G=E^{ab}J_{a}\p_{b}-{1\over 6}f^{abc}\p_{a}\p_{b}\p_{c}\eqno(23)$$
where
$$E={1\over k}\left(\matrix{1&0&0&0\cr 0&1&0&0\cr 0&0&0&1\cr
0&0&1&-b+1/2k\cr}\right)\eqno(24)$$
and the group indices are always raised and lowered with the
invariant metric $G_{ab}$.
The only non-zero component of $f^{abc}$ is $f^{124}=1/k^2$.
It can be verified that (24) satisfies the superconformal ME
\cite{GHKO}
and preserves the current algebra structure.
In particular
$$G(z)G(w)={2c/3\over (z-w)^3}+{2T(w)\over (z-w)}+\rr\eqno(25)$$
with $c=6$, $T=T^{b}+T^{f}$, with $T^{b}$ being the affine Sugawara
stress tensor and $T^{f}$ the free fermionic stress tensor
$$T^{f}=-{1\over 2}G^{ab}\p_{a}\d \p_{b}\eqno(26)$$
$G$ is also primary with conformal weight $3/2$ with respect to $T$.
The supersymmetric currents in the theory are the superpartners of
the free
fermions
$$G(z)\p_{a}(w)={\hat J_{a}(w)\over (z-w)}+\rr\eqno(27)$$
It is easy to find that
$$\hat P_{1}=P_{1}-{1\over k}\p_{2}\p_{T}\;\;,\;\;\hat
P_{2}=P_{2}+{1\over k}\p_{1}\p_{T}\eqno(28)$$
$$\hat J=J+{1\over 2k}T-{1\over k}\p_{1}\p_{2}\;\;,\;\;\hat
T=T\eqno(29)$$
The supersymmetric currents $\hat J_{a}$ satisfy the same algebra
(1) as the bosonic ones $J_{a}$. This should be contrasted with the
compact case
where the level is shifted by the dual Coxeter number.
The supercurrent $G$ has a similar expression in terms of $\hat
J_{a}$ as in the compact case \cite{KS}
$$G=G^{ab}\hat J_{a}\p_{b}-{1\over
3}f^{abc}\p_{a}\p_{b}\p_{c}\eqno(30)$$

As in the bosonic case \cite{kk} the supersymmetric current algebra
can be written in terms of free bosons, $x^{\pm},x,\bar x$ and
fermions
$\p^{\pm},\p,\pp$ with
$$<x^{+}(z)x^{-}(w)>=-<x(z)\bar x(w)>=2\log(z-w)\eqno(31)$$
$$<\p(z)\pp(w)>=-<\p^{+}(z)\p^{-}(w)>={2\over (z-w)}\eqno(32)$$
all others being zero.
Explicitly,
$$P_{1}+iP_{2}=i\sqrt{k}e^{-ix^{-}/Q}\d
x\;\;,\;\;P_{1}-iP_{2}=i\sqrt{k}e^{ix^{-}/Q}\d \bar x\eqno(33)$$
$$J={Q\over 2}\d x^{+}+{1\over 2}\left(Q+{1\over Q}\right)\d x^{-}
+{i\over 2}\p\pp\;\;,\;\;T={Q\over b}\d x^{-}\eqno(34)$$
$$\p_{1}+i\p_{2}=\sqrt{k}e^{-ix^{-}/Q}\p\;\;,\;\;\p_{1}-i\p_{2}=
\sqrt{k}e^{ix^{-}/Q}\pp\eqno(35)$$
$$\p_{J}=i{Q\over
2}\left(\p^{+}+\p^{-}\right)\;\;,\;\;\p_{T}=i{Q\over b}\p^{-}
\eqno(36)$$
where $Q=\sqrt{kb}$.
The fermionic admixture to $J$ is added to guarantee that $J_{a}$ and
$\p_{a}$
commute.

It is not difficult to see that the stress tensor and supercurrent
are
free in the free-field basis
$$T={1\over 2}\left[\d x^{+}\d x^{-}-\d x \d \bar x\right]+{1\over
4}\left[
\p^{+}\d \p^{-}+\p^{-}\d \p^{+}-\p\d \pp-\pp\d\p\right]\eqno(37)$$
$$G={i\over 2}\left[ \d x\pp+\d \bar x\p+\d x^{-}\p^{+}+\d
x^{+}\p^{-}\right]
\eqno(38)$$
In the free field basis it can be easily seen that the theory has an
N=2 superconformal algebra. The conventionally normalized U(1)
current that determines the "complex
structure" is given by
$$\j={1\over 2}\left[\p\pp-\p^{+}\p^{-}\right]\eqno(39)$$
whereas the second supercurrent is
$$G^{2}=-{1\over 2}\left[\d x \pp-\d \bar x\p+\d x^{+}\p^{-}-\d
x^{-}\p^{+}\right]\eqno(40)$$

In conformity with our $\s$-model discussion later , we will also
consider
the presence of the dilaton, which here is reflected as background
charge
in the lightcone directions.
We have the following modifications for the N=2 generators
$$\delta\j=-i\left(Q^{+}\d x^{-}+Q^{-}\d
x^{+}\right]\;\;\;,\;\;\;\delta T={i\over
2}\left(Q^{+}\d^{2}x^{-}-Q^{-}\d^{2}x^{+}\right)\eqno(41)$$
$$\delta G^{1}=Q^{+}\d \p^{-}-Q^{-}\d \p^{+}\;\;\;,\;\;\;\delta
G^{2}=i\left(Q^{+}\d \p^{-}+Q^{-}\d \p^{+}\right)\eqno(42)$$
and $\delta c=-8Q^{+}Q^{-}$.

The final step is the realization that when $Q^{-}=0$ the theory has
even larger superconformal symmetry, namely N=4.
The N=4 superconformal algebra in question contains the stress
tensor, $SU(2)_{k}$ currents $J^{a}$  and four supercurrents that
transform as conjugate doublets under the $SU(2)$.
It is defined in terms of the OPEs
$$J^{a}(z)J^{b}(w)={k\over 2}{\delta^{ab}\over
(z-w)^2}+i\epsilon^{abc}{J^{c}(w)\over (z-w)}+\rr\eqno(43)$$
$$J^{a}(z)G^{i}(w)={1\over 2}\s^{a}_{ji}{G^{j}(w)\over
(z-w)}+\rr\;\;,\;\;
J^{a}(z)\bar G^{i}(w)=-{1\over 2}\s^{a}_{ij}{\bar G^{j}(w)\over
(z-w)}+\rr
\eqno(44)$$
$$G^{i}(z)\bar G^{j}(w)={4k\delta^{ij}\over
(z-w)^3}+2\s^{a}_{ji}\left[
{2J^{a}(w)\over (z-w)^2}+{\d J^{a}(w)\over
(z-w)}\right]+2\delta^{ij}{
T(w)\over (z-w)}+\rr\eqno(45)$$
the rest being regular.
$J^{a}$, $G^{i},\bar G^{i}$ are primary with the appropriate
conformal weight
and $c=6k$.
In our case the SU(2) current algebra has level $k=1$.
The N=4 generators, in the free field basis are
$$G^{1}={i\over \sqrt{2}}\left[\d\bar x\p+\d
x^{-}\p^{+}\right]\;\;,\;\;
G^{2}=-{i\over \sqrt{2}}\left[\d \bar x\p^{-}+\d x^{-}\pp\right]
e^{iQ^{+}x^{-}}\eqno(46)$$
$$\bar G^{1}={i\over \sqrt{2}}\left[\d x \pp+\d
x^{+}\p^{-}-2iQ^{+}\d\p^{-}\right],\bar G^{2}={i\over \sqrt{2}}\left[
\d x \p^{+}+\d x^{+}\p
-iQ^{+}\p\p^{+}\p^{-}\right]e^{-iQ^{+}x^{-}}\eqno(47)$$
$$J^{1}+iJ^{2}={1\over 2}\p\p^{+}e^{-iQ^{+}x^{-}}\;\;,\;\;
J^{1}-iJ^{2}={1\over 2}\pp\p^{-}e^{iQ^{+}x^{-}}\eqno(48)$$
$$J^{3}={1\over 4}\left[\p\pp-\p^{+}\p^{-}-2iQ^{+}\d
x^{-}\right]\eqno(49)$$
and $T$ is given in (37).

One can invert the map to free fields and right these operators in
terms of the original ($\s$-model) currents. Care should be exercized
though in normal ordered expressions.
What we need here is the appearence, in the N=4 generators, of terms
which are non-local in the original currents.
These terms are of the form $exp\left[\pm{i\over k}(QQ^{+}-1)\int T
dz\right]$.
However, in the sigma model, the current $T$ is given as a total
derivative of
a free field, $T=\d u$, with $\d\db u=0$.
Thus the exponential factors are just $exp\left[\pm{i\over
k}(QQ^{+}-1)u(z)\right]$.
Notice also that they dissapear at a specific value of the background
charge, $Q^{+}=1/Q$.

We have seen above that the $E^{c}_{2}$ gravitational  wave
background
has a large (N=4) superconformal symmetry on the worldsheet. We will
subsequently show that it belongs in fact to the $N=4$ class of
solutions described in \cite{kkl}, and thus the associated string
theory has an N=4 spacetime supersymmetry (in the type II case).
We will start in 4-d Euclidean space by introducing two complex
coordinates
$u$ and $v$.
Then
the dilaton field is a function of
$u,\bar u,v,\bar v$: $\Phi=\Phi(u,\bar u,v,\bar v)$.
To describe the four-dimensional metric and
non-constant antisymmetric tensor field background
we will use the $N=2$ superfield formalism of \cite{ghr} and refer
for
details
to our general analyis \cite{kkl} on supersymmetric backgrounds in
string theory. Specifically, we introduce one  $N=2$
chiral superfield $U$ with $u$ as complex bosonic coordinate
and one twisted chiral superfield $V$ with $v$ as complex bosonic
coordinate.
It follows that
 the metric and antisymmetric tensor field background can be also
entirely
derived from a single real function $K(U,\bar U,V,\bar V)$, the
socalled quasi K\"ahler potential:
$$
G_{\mu\nu}=\pmatrix{0&K_{u\bar u}&0&0\cr K_{u\bar u}&0&0&0
\cr 0&0&0&-K_{v\bar v}\cr 0&0&-K_{v\bar v}&0\cr},\qquad
B_{\mu\nu}=\pmatrix{0&0&0&K_{u_i\bar v_p}\cr
0&0&K_{v_p\bar u_i}&0
\cr 0&-K_{v_p\bar u_i}&0&0\cr -K_{u_i\bar v_p}&0&0&0\cr},
\eqno(50)
$$
where $K_{u\bar u}={
\partial^2 K\over\partial U\partial\bar U}$, etc. The field
strength $H_{\mu\nu\lambda}$ can also be expressed in terms
of the function $K$ as $H_{u\bar uv}={\partial^3 K\over \partial
U\partial
\bar U\partial V}$, etc.

The field equations, i.e. the the conditions of the vanishing of
the (one-loop) $\beta$-functions will now lead to some partial
differential equations for the two functions $K$ and $\Phi$
\cite{kkl}.
Before we show that the gravitational wave background is (after some
analytic
continuation) a solution of these equations, we would first present a
general class of solutions with N=4 superconformal symmetry.
The condition for N=4 superconformal symmetry is \cite{ghr}
$$
\Kuu=e^\lambda\Kvv,\qquad\lambda={\rm const.}\eqno(51)
$$
and $\delta c=0$.
As derived in \cite{kkl}, the vanishing of the metric and
antisymmetric
tensor field $\beta$-functions leads to the following
set of equations:
$$\partial_u(\U-2\Phi)=\bar C_1(\bar u)\Kuu\;\;,\;\;
\partial_v(\U-2\Phi)=\bar C_2(\bar v)\Kuu \eqno(52 a)$$
$$\partial_{\bar u}(\U-2\Phi)=C_1(u)\Kuu\;\;,\;\;
\partial_{\bar v}(\U-2\Phi)=C_2(v)\Kuu\eqno(52 b)
$$
and
$$\eqalign{
&\partial_u\partial_{\bar
u}(\U-2\Phi)=e^\lambda\partial_v\U\partial_{\bar v}
(\U-2\Phi),\cr
&\partial_v\partial_{\bar v}(\U-2\Phi)=e^{-\lambda}\partial_u\U
\partial_{\bar u}(\U-2\Phi)\cr}\eqno( 53 a)$$
$$\partial_u\Phi\partial_{\bar u}\U=\partial_{\bar
u}\Phi\partial_u\U\;\;,\;\;
\partial_v\Phi\partial_{\bar v}\U=\partial_{\bar
v}\partial_v\U.\eqno(53b)
$$
The central charge deficit, related to the dilaton $\beta$-function
becomes
$$
\eqalign{
\delta c=3\alpha' {1\over\Kuu}\lbrace
8\partial_{u}\Phi
\partial_{\bar u}\Phi-2\partial_{u}\partial_{\bar u}\U
-4\partial_u\U\partial_{\bar u}\U\cr
-e^{-\lambda}
(8\partial_{v}\Phi
\partial_{\bar v}\Phi-2\partial_{v}\partial_{\bar v}\U
-4\partial_v\U\partial_{\bar v}\U)\rbrace .\cr}\eqno(54)
$$
{}From eqs.(52) and (53) we obtain after some algebra that
$$
\eqalign{&C_1(u)=Cu+c_1,\qquad \bar C_1(\bar u)=C\bar u +\bar c_1,\cr
&C_2(v)=-e^{-\lambda}Cv+c_2,\qquad \bar C_2(\bar
v)=-e^{-\lambda}C\bar
v+\bar
c_2.\cr}\eqno(55)
$$
where $C,c_{1},\bar c_{1},c_{2},\bar c_{2}$ are constants.
If $C\neq 0$, $c_1,c_2,\bar c_1,\bar c_2$ can be set to
zero
by simple translations in the coordinates. Then equations (52,53)
possess the following unique solution:
$$\Kuu={a\over u
\bar u-e^{-\lambda}v\bar v}\;\;,\;\;
	\Phi=-{1\over 2}(1+Ca)\log(u\bar u
-e^{-\lambda}v\bar v).\eqno(56)
$$
with $ a$ constant.
For $\lambda=i\pi$ and $a=1$, this background describes the so called
semi wormhole space \cite{chs,koun}. The corresponding quasi
K\"ahler
function can be found in \cite{kkl}. This background
corresponds to the exact conformal field theory, based on the WZW
model
$SU(2)\times U(1)$. The central charge deficit $\delta c$ is zero
by choosing an appropriate value for the background charge of the
$U(1)$.
Then one can explicitely construct an $N=4$ superconformal algebra
from the currents of the WZW model \cite{koun}.

{}From now on we assume that $C=0$.

\noindent ({\bf i}): If $c_1=c_2=\bar c_1=\bar c_2=0$ we deal with
the
case
already discussed in ref.\cite{kkl}:
$$\U-2\Phi={\rm const}.\eqno(57)$$
This class
of solutions describes the socalled axionic instanton background with
$\delta c=0$.

\noindent ({\bf ii}): If $c_1=0$, $c_2,\bar c_1,\bar c_2\neq 0$ or
 $c_1=\bar c_1=0$, $c_2,\bar c_2\neq 0$ it follows that
the metric and antisymmetric tensor field backgrounds
are trivial: $U={\rm const}$.
 The dilaton field may be linear in some of the coordinates.

\noindent ({\bf iii}): If $c_1, c_2,\bar c_1,\bar c_2\neq 0$
eqs(52,53)
imply that $\Kuu=\Kuu\biggl({u\over c_1}+{\bar u\over \bar
c_1}+e^{-\lambda}(
{v\over c_2}+{\bar v\over\bar c_2 })\biggr)$
and $\Phi=\Phi\biggl({u\over c_1}+{\bar u
\over\bar c_1}+e^{-\lambda}({v\over c_2}+{\bar v\over\bar
c_2})\biggr)$
with,
in addition, $2\Phi'={\Kuu'\over \Kuu}-c_1\bar c_1\Kuu$.
Assuming $\Phi={\rm const}$, we obtain the following expression for
the
background metric $(c_1=c_2=\bar c_1=\bar c_2=1$):
$$K_{u\bar u}=-{1\over u+\bar u+e^{-\lambda}(v+\bar v)}\eqno(58)
$$

\noindent ({\bf iv}): $c_1=c_2=0$, $\bar c_1,\bar c_2\neq 0$.
This case can be treated similarly to the pevious case (iii) and
leads
to $\Kuu=\Kuu\biggl({u\over \bar c_2}+{v\over \bar c_1}\biggr)$.

\noindent ({\bf v}): $\bar c_1=c_2=\bar c_2=0$, $c_1\neq 0$.
Then the field equations are uniquely solved by the following
background:
$$\Phi={\rm const}, \qquad \Kuu={a\over u},\eqno(59)
$$
leading to $\delta c=0$. The corresponding metric background is
described by
the following line element:
$${\rm d}s^2={a\over u}\biggl({\rm d}u{\rm d}\bar u-e^\lambda{\rm
d}v{\rm d}
\bar v\biggr).\eqno(60)
$$
This solution follows from the following quasi K\"ahler function,
$$K=a\biggl(\bar u\log u
+e^\lambda{v\bar v\over u}\biggr),\eqno(61)
$$
and the antisymmetric tensor field is given as
$B_{u\bar v}=ae^\lambda v/ u^2$.

As we will show now, this supersymmetric background is just the
gravitational wave background for $\lambda=i\pi$ and
$a=-{1\over 2}$, performing
appropriate changes of coordinates and analytic continuations.
Explicitly, we go to the Minkowskian by treating $u$ and $\bar u=w$
as
independent  coordinates. Then changing the
 coordinates as $u\rightarrow -{1\over 2}\log(-2u)$, $w\rightarrow w
+x^2$, $\theta\rightarrow\theta +{u\over 2}$ ($v=xe^{i\theta}$) we
obtain the following four-dimensional metric from the conformally
flat
metric eq.(60):
$${\rm d}s^2={\rm d}x^2+x^2{\rm d}\theta^2+x^2{\rm d}u^2+{\rm d}u{\rm
d}w.
\eqno(62)
$$
The antisymmetric tensor becomes
$B_{u\theta}=2x^2$.
Finally, substituting $u\rightarrow iu$, $w\rightarrow iw$ and
setting $x\cos\theta=x_1+\cos u x_2$, $x\sin\theta=\sin u x_2$
and $w\rightarrow w+{1\over 2}x_1x_2\sin u$, the background
exactly agrees with the gravitational wave background as determined
by the
$\sigma$-model action (6).
However the list of solutions above still share the property that
they have N=4 spacetime supersymmetry and the presence of a null
Killing symmetry. They can be viewed as generalizations of the
background studied
in this paper.

There are several associated problems which remain open and whose
answer will illuminate our present understanding of the physics of
stringy gravity.
One immediate problem is to finish the calculation of scattering
amplitudes
in the gravitational wave background. So far we have been unable to
calculate
scattering amplitudes in any non-trivial space-time background (i.e.
non-compact)
of string theory, and we think that such a task will provide some
interesting insights into scattering matrices in curved spaces.
Eventually we need to understand the GSO projection in the associated
superstring  theory in order to obtain a tachyon free spectrum.
Finally it would be interesting to examine the possibility that such
wave profiles could be observed and if they have any characteristic
stringy signature which might be measurable.

\vskip 2cm
\noindent
{\bf Acknowledgments}

The work of C. Kounnas was partially supported by EEC grants
SC1$^{*}$-0394C and
SC1$^{*}$-CT92-0789.\\


\noindent

\newpage



\begin{thebibliography}{6666}
\bibitem{gw} D. Amati and C. Klim\v c\'\i k, Phys. Lett. {\bf B219}
(1989) 443;\\
R. G\"uven, Phys. Lett. {\bf B191} (1987) 275;\\
H. de Vega and N. Sanchez, Nucl. Phys. {\bf B317} (1989) 709;\\
G. T. Horowitz and A. R. Steif, Phys. Rev. Lett. {\bf 64} (1990) 260;
Phys. Rev. {\bf D42} (1990) 1950; A. Steif, Phys. Rev. {\bf D42}
(1990)
2150;\\
A. A. Tseytlin, Phys. Lett. {\bf 288} (1992) 279; Nucl. Phys. {\bf
B390} (1993) 153; Phys. Rev. {\bf D47} (1993) 3421.

\bibitem{w} C. Nappi and E. Witten, IAS preprint, hepth/9310112.

\bibitem{kk} E. Kiritsis and C. Kounnas, Phys. Lett. {\bf B320}
(1994)
264.

\bibitem{kt} C. Klimcik and A. Tseytlin, hep-th/9311012,9402120.

\bibitem{sf} K. Sfetsos, hep-th/9311010,9311093,9402031.

\bibitem{osr} D. Olive, E. Rabinovici, A. Schwimmer, Phys.Lett. {\bf
B321}
(1994) 361.

\bibitem{r1} O. Jofre and C. Nunez, hep-th/9312189.

\bibitem{rest} N. Mohammedi, hep-th/9312182;\\
A. Kumar and S. Mahapatra, hep-th/9301098;\\
J. M. Figueroa-O'Farrill and S.Stanciu, hep-th/9402035;\\
A. Kehagias and P. Maessen, hep-th/9403041;\\
I. Antoniadis and N. Obers, hep-th/9403191;\\
K. Sfetsos and A. Tseytlin, hep-th/9404063.



\bibitem{kkl} E. Kiritsis, C. Kounnas and D. L\"ust, Int. J. Mod.
Phys.
{\bf A9} (1994) 1361.

\bibitem{HK} M. B. Halpern and E. Kiritsis, Mod. Phys. Lett. {\bf A4}
(1989) 1373;\\
A. Morozov, A. Perelomov, A. Rosly, M. Shifman and A. Turbiner, Int.
J.
Mod. Phys. {\bf A5} (1990) 803.

\bibitem{str} R. Streater, Comm. Math. Phys. {\bf 4} (1967) 217.

\bibitem{sf2} K. Sfetsos, Phys.Lett. {\bf B271} (1991) 301.

\bibitem{BK} I. Bakas and E. Kiritsis, Int. J. Mod. Phys. {\bf A7}
[Supp. A1] (1992) 55.

\bibitem{GHKO} A. Giveon, M. Halpern, E. Kiritsis, N. Obers, Int. J.
Mod. Phys. {\bf A7} (1992) 947.


\bibitem{KS} P. Di Vecchia, V. Knizhnik, J. Petersen and P. Rossi,
Nucl. Phys. {\bf B253} (1985) 701.

\bibitem{ghr} S. Gates, C. Hull and M. Rocek, Nucl. Phys. {\bf B248}
(1984) 157.

\bibitem{chs} C. Callan, J. Harvey and A. Strominger, Nucl. Phys.
{\bf
B359} (1991) 611.

\bibitem{koun} C. Kounnas, Phys. Lett. {\bf B321} (1994) 26;
hepth/9302058,9402080;\\
I. Antoniadis, S. Ferrara and C. Kounnas, hep-th/9402073.


\end{thebibliography}
\end{document}